\documentstyle[12pt,twoside,fleqn,espcrc1_m]{article}

\input psfig.sty
\def\lsim{\mathrel{\rlap{\lower4pt\hbox{\hskip1pt$\sim$}}
    \raise1pt\hbox{$<$}}}         %less than or approx. symbol
\def\gsim{\mathrel{\rlap{\lower4pt\hbox{\hskip1pt$\sim$}}
    \raise1pt\hbox{$>$}}}         %greater than or approx. symbol
\newcommand{\AmS}{{\protect\the\textfont2
  A\kern-.1667em\lower.5ex\hbox{M}\kern-.125emS}}

\title{Fundamental Symmetries and Theory}

\author{W. C. Haxton\address{
Institute for Nuclear Theory, Box 351550,
and Department of Physics, Box 351560, \\
University of Washington, Seattle, Washington  98195}}

\begin{document}
% typeset front matter
\maketitle

\begin{abstract}
Nuclei are powerful laboratories for studying fundamental symmetries
because they filter and enhance specific interactions.
I discuss four examples --- hadronic parity violation, atomic
electric dipole moments, precision $\beta$ decay tests, and 
nuclear tests of neutrino masses --- to illustrate some of the 
progress that has been made in the past few years.
\end{abstract}

\section{Introduction}
Let me begin by thanking Bernard Frois and the Organizing Committee
for making it possible for us to enjoy both the 1998 International
Nuclear Physics Conference and the lovely city of Paris.
My charge today is to summarize fundamental symmetries and theory,
a topic with a rich history in nuclear physics.  
A generation ago nuclear studies of $\beta$ decay helped 
established the V-A nature of the weak interaction, the 
conserved vector current hypothesis, and other aspects of the
weak interaction that became part of the standard model's experimental foundation.
Much of the work carried on today is in searches for cracks in that
foundation, subtle violations of low-energy symmetries that may
indicate the nature of physics beyond the standard model.
The field has become much too broad to cover adequately in one
talk.  Thus the best I can do this morning is to try to capture
the flavor of the field by presenting four selected vignettes.
Two of these --- the discussions of hadronic parity violation 
and precision $\beta$ decay experiments --- focus primarily on the standard model.
Two others --- time reversal tests and massive neutrinos ---
look beyond the standard model to the new physics
that we expect to characterize the next level of unification.

\section{Nuclear Parity Violation}
The hadronic weak neutral current is relatively difficult to 
isolate.  Although the weak interaction mediates flavor-changing
decays of the $\Lambda$ and other strange baryons, there are
no tree-level flavor-changing neutral currents in the standard
model.  Furthermore, weak radiative corrections to $Z^0$ exchange,
while flavor changing, are GIM suppressed.  Thus to see
the hadronic weak neutral current one must study $\Delta$ S = 0
interactions, which effectively limits one to NN interactions
and nuclei.  As the much stronger strong and electromagnetic
interactions also contribute to NN interactions, the parity 
violation of the weak interaction must be exploited as the
experimental filter \cite{adel}.  

At low energies the parity violating NN interaction can be 
modeled as a meson exchange interaction, in analogy with 
similar descriptions of the strong interaction.  One
meson-nucleon vertex contains the weak interaction, while the
second is strong, as depicted in Fig. 1. 
  
\begin{figure}[htb]
\psfig{bbllx=0.0cm,bblly=8.0cm,bburx=18cm,bbury=15.0cm,figure=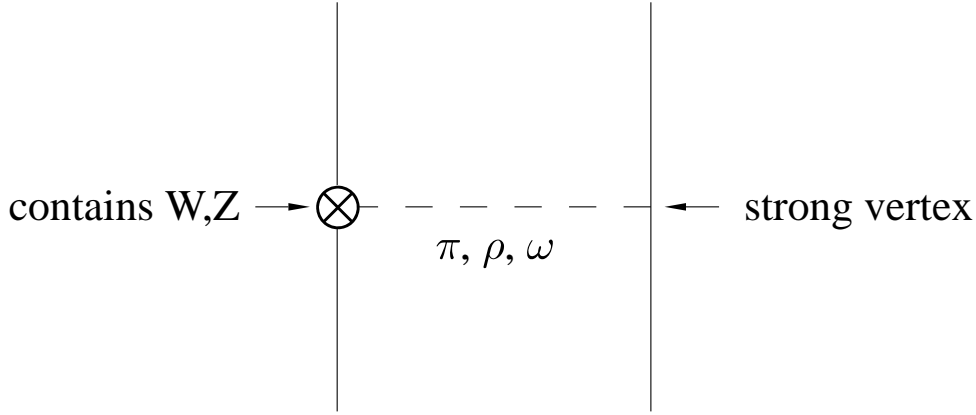,height=2.0in}
\caption{One-boson-exchange potential for the PNC NN interaction.
One vertex is weak, containing the W or Z exchange, and one strong.}
\end{figure}
  
In the long-wavelength limit such a description in terms of 
$\pi$, $\rho$, and $\omega$ exchanges is sufficiently general
to produce all five independent weak s-p amplitudes.  To the
extent that the range of the weak NN interaction is reasonably
represented by such a description, it also provides a model
for p-d and higher partial wave interactions. 
  
In the standard model the low-energy hadronic weak interaction is
a product of charged and neutral currents
\[ L^{eff} = {G \over \sqrt{2}} \left[ J_W^\dagger J_W + J_Z^\dagger J_Z \right] + h.c. \]
where the charged current is comprised of $\Delta$S = 0 
$\Delta$I = 1 and $\Delta$S = -1 $\Delta$I = 1/2
pieces,
\[ J_W = \mathrm{cos} \theta_c J_W^{\Delta S = 0~\Delta I = 1} + \mathrm{sin} \theta_c J_W^{\Delta S = -1~\Delta I = 1/2}. \]
Thus the effective Lagrangian for $\Delta$S = 0 NN interactions is
\[ L^{eff}_{\Delta S = 0} = {G \over \sqrt{2}}
\left[ \mathrm{cos}^2 \theta_c J_W^{0 \dagger} J_W^0 + \mathrm{sin}^2 \theta_c J_W^
{1 \dagger} J_W^1 + J_Z^{\dagger} J_Z \right]. \]
Note that the first term above, a symmetric combination of 
$\Delta$ I = 1 currents, has $\Delta$ I = 0 and 2, only.  The
second term, a symmetric combination of $\Delta$ I =1/2 currents,
has $\Delta$ I = 1 but is suppressed by the square of the 
Cabibbo angle.  Thus we conclude that the $\Delta$ I = 1 weak
NN interaction should be dominated by the third (neutral current)
term above.  In meson exchange models this is the channel 
dominated by $\pi^\pm$ exchange, the longest range component
of the NN interaction.  We conclude that the weak $\pi$NN coupling
$f_\pi$ contains the neutral current interaction we are seeking.

This coupling is the natural meeting point between experiment
and nuclear theory on one hand, and the standard model on the
other.  As in the work of Desplanques, Donoghue, and Holstein (DDH) \cite{ddh},
$f_\pi$ can be estimated from standard model bare quark couplings
plus calculated strong interaction dressings: calculations of the
latter are clearly quite difficult and model dependent, and
could yield effective $\pi$NN couplings that are quite different
from the underlying bare couplings.  The goal is to compare
the resulting $f_\pi$ to experiment, thereby testing whether 
our theory techniques are adequate for calculating the 
strong interaction corrections. 

As there are five elementary s-p amplitudes, the simplest
experimental strategy would be to make five independent NN
measurements to separate out $f_\pi$.  Unfortunately only one
quantity has been determined experimentally,
\[A_L(pp) = {\sigma^+(\theta) - \sigma^-(\theta) \over
\sigma^+(\theta) + \sigma^-(\theta) } \sim 2 \cdot 10^{-7} \]
This has meant that the needed additional constraints must be
taken from nuclear experiments, where nuclear structure 
uncertainties can make it difficult to extract weak
meson-nucleon coupling constants reliably.  Fortunately there
are several cases (e.g., $^{18}$F and $^{19}$F) where these
uncertainties can be largely eliminated through ancillary
experiments.  

Nuclei also offer some compensating advantages: 
The mixing of states of definite isospin provides an
isospin filter that can be exploited to isolate specific 
couplings, such as $f_\pi$.
Furthermore, there are attractive opportunities for using nuclear
degeneracies to greatly enhance the parity violating signal. 
  
These properties are nicely illustrated by the example of $^{18}$F.
The level structure and the effects of the parity violation are
illustrated in Fig. 2.  The observable is the circular
polarization of the photons emitted in the decay of the 1081 keV
$0^-0$ state.  Because the $0^+1$ 1042 keV state is only 39 keV
away, almost all of the parity violation is attributable to the
mixing of these two states.  This has the nice consequence that
this observable tests the $\Delta$I = 1 part of $V^{PNC}$. 

\begin{figure}[htb]
\psfig{bbllx=-2.0cm,bblly=4.0cm,bburx=18cm,bbury=12.3cm,figure=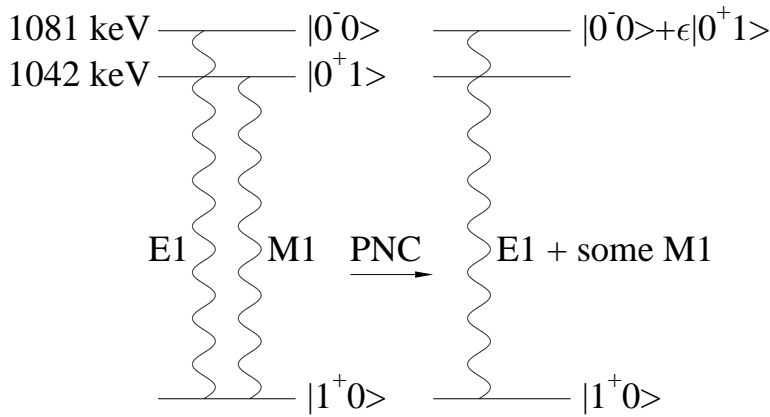,height=2.0in}
\caption{Schematic diagram showing the effects on parity mixing
on the decay of the 1081 keV state in $^{18}$F.}
\end{figure}
  
The enhancement comes both from the small energy denominator 
for the two-state mixing, and from the strength of the parity-odd
M1 multipole that the parity admixture introduces into the 
decay of the 1081 keV state.  The parity-allowed E1 decay of 
the $0^-0$ state is very weak: the long-wavelength limit of the 
isoscalar E1 operator cannot generate transitions.  The admixed
M1 is unusually strong, on the order of 10 W.u.  Thus
\[P_{\gamma} \sim 2 Re \left[ {<1^+ \parallel M1 \parallel 0^+> \over
<1^+ \parallel E1 \parallel 0^->} {<0^+1 \parallel V_{PNC}^{\Delta I = 1}
\parallel 0^-0> \over \Delta E} \right] \sim 10^{-3}  \]
The $M1/E1$ matrix element ratio is $\sim$ 110, while
$\langle V \rangle/\Delta E \sim 10^{-5}$, compared to the
natural scale of weak interactions,
\[  {4 \pi G^2 m_\pi^2 \over 
g_{\pi NN}^2 } \sim 10^{-7} . \]
It also turns out the the nuclear matrix element can be 
effectively measured from the axial-charge $\beta$ decay of 
$^{18}$Ne, as described in \cite{adel}. 

The $^{18}$F result --- actually an upper bound --- is one of 
several measurements in the NN, few-body, and light nuclear
systems which can be interpretted, i.e., where the nuclear 
physics uncertainties are sufficiently under control that 
the weak meson-nucleon couplings can be reliably extracted:
\[ \left. \begin{array}{ll}
A_L(\vec{p}+p) & 15,45 ~\mathrm{MeV} \\
A_L(\vec{p}+^4He) & 46 ~\mathrm{MeV} \\
A_L(\vec{p}+d) & 15 ~\mathrm{MeV} ~\mathrm{(upper bound)} \\
P_\gamma(^{18}F) & \\
A_\gamma(^{19}F) & 
\end{array} \right\} \Rightarrow  
h^0_{\rho}+0.5h^0_{\omega} \sim 1.2 (h^0_{\rho}+0.5h^0_{\omega})^{DDH},~~f_{\pi}
\sim 0 \]
Here $f_\pi$ is the isovector weak $\pi$NN coupling and 
$(h^0_\rho+0.5h^0_\omega)$ is the combination of $\rho$ and 
$\omega$ isoscalar weak couplings that arises in
calculations of the $^{19}$F and $\vec{p}+^4$He asymmetries.
The superscript DDH denotes the best value of Ref. \cite{ddh}.
The net result is a surprise: although the isoscalar combination
is about as predicted by DDH, $f_\pi$ is consistent with zero
and no larger than about one-third $f_\pi^{DDH}$. 
   
So a summary of things as of a year ago would be: \\
$\bullet$ No evidence has been found for the neutral hadronic current. \\ 
$\bullet$ The isospin anomaly, $f^{PNC}_{\pi} \lsim {1 \over 3} f^{DDH}_
{\pi}$ while the isoscalar combination has the expected size,
is superficially similar to the $\Delta$I = 1/2 rule in strangeness-changing
decays: the isospin dependence of the meson-nucleon couplings 
differs significantly from those of the underlying bare couplings.
Of course, in the case of the $\Delta$I = 1/2 rule, it is an
enhancement in the $\Delta$I = 1/2 amplitudes. \\
$\bullet$ Presumably the explanation for both effects must be 
connected with some interesting strong interaction physics 
occurring in the weak vertices. \\
  
A new constraint was obtained recently from a novel measurement,
the first detection of a nuclear anapole moment.  Atomic PNC experiments
are sensitive to a variety of new physics, while also testing standard
model couplings to high precision.  The dominate standard model interaction is
\[ A(e) - V(N). \]
That is, the leading PNC amplitude involves an axial coupling
of the exchanged $Z^0$ to the electron and a coherent vector
coupling to the nucleus.  The nuclear weak charge is approximately
its neutron number.  About a year ago the Boulder group \cite{boulder} announced
an atomic PNC result of unprecedented sensitivity, a $\sim$ 0.3\% measurement
in $^{133}$Cs.  This measurement, for example, limits the scale
of new $Z'$s to lie above 1.4 TeV --- a value limited by the
associated atomic theory, which is believed to be accurate to $\sim$ 1.0\%. 
  
\begin{figure}[htb]
\psfig{bbllx=1.0cm,bblly=7.6cm,bburx=18cm,bbury=14.7cm,figure=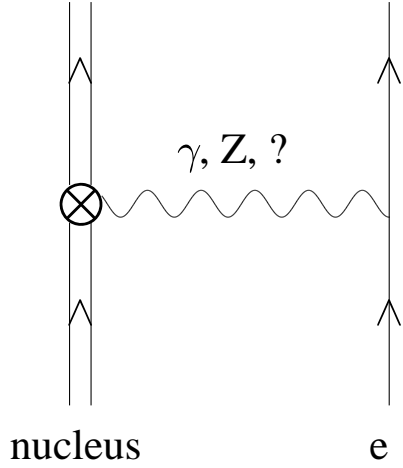,height=2.0in}
\caption{Atomic interactions are mediated by $\gamma$ and $Z^0$
exchange, as well as new physics effects, denoted by ?,
associated with extra $Z$s, leptoquarks, compositeness, etc.}
\end{figure}
  
The connection to nuclear PNC is the contribution associated with
the axial coupling to the nucleus.  There is a weak tree-level
contribution to PNC from $Z^0$ exchange of the form
\[ V(e) - A(N). \]
Surprisingly, even more important in a heavy nucleus are contributions due to
radiative corrections.  Recall that the possible static 
electromagnetic couplings to the nucleus are
\[ \begin{array}{cccc}
&\underline{CJ}&\underline{MJ}&\underline{EJ} \\
J=0&PT&& \\
J=1 & \not P \not T & PT & \not P T \\
J=2 & PT & \not P \not T & P \not T \\
J=3 & \not P \not T & PT & \not P T \\
\vdots & \vdots & \vdots & \vdots
\end{array} \]
For example, the electromagnetic current matrix element for a 
spin-1/2 particle like the nucleon involves the first four entries in the above table
\[ \begin{array}{c} <g.s. \parallel J_\mu^{em} \parallel g.s.> =
\bar{N}(p') (\overbrace{F_1 \gamma_\mu}^{C0}+
\overbrace{F_2 \sigma_{\mu \nu} q^{\nu}}^{M1}+ 
\underbrace{F_A \gamma_5 \gamma_\mu + F_P \gamma_5 q_\mu}_
{E1: \not P T anapole}+\underbrace{F_T \gamma_5
\sigma_{\mu \nu} q^\nu}_{C1: \not P \not T edm} ) N(p) \end{array} \]
where current conservation demands $F_A = {q^2 \over 2M} F_P$.
This PNC term thus can be written
\[ <p' \parallel J_\mu^{em} \parallel p>_{\not P T} =
{a(q^2) \over M^2} \bar{N}(p') (\not q q_\mu - q^2 \gamma_\mu)
\gamma_5 N(p) \]
where $a(q^2)$ is the anapole moment,
first discussed by Zeldovich \cite{zeld,khrip}.  It transformations under rotations
like the nucleon spin.  It vanishes for on-shell photons, but for
virtual photons generates a contact interaction between the scattered 
electron and the nucleon.

Figure 4a shows one contribution to the nucleon anapole moment:
it arises as a weak radiative correction.
Fig. 4b shows a similar weak radiative correction that cannot be
written as a nucleon electromagnetic moment.  This shows that the
anapole moment is not really a measurable (i.e., gauge invariant)
quantity, though for a nucleus the dominant contribution to the
anapole moment is separately gauge invariant.  Fig. 4c is the
usual tree-level V(e)-A(N) $Z^0$ exchange, where
the axial nuclear coupling is isovector, which combines with the
electron--nuclear anapole interaction (Fig. 4d) to give the nuclear-spin-dependent
parity violation in an atom.
  
\begin{figure}[htb]
\psfig{bbllx=-1.0cm,bblly=3.0cm,bburx=16cm,bbury=21.9cm,figure=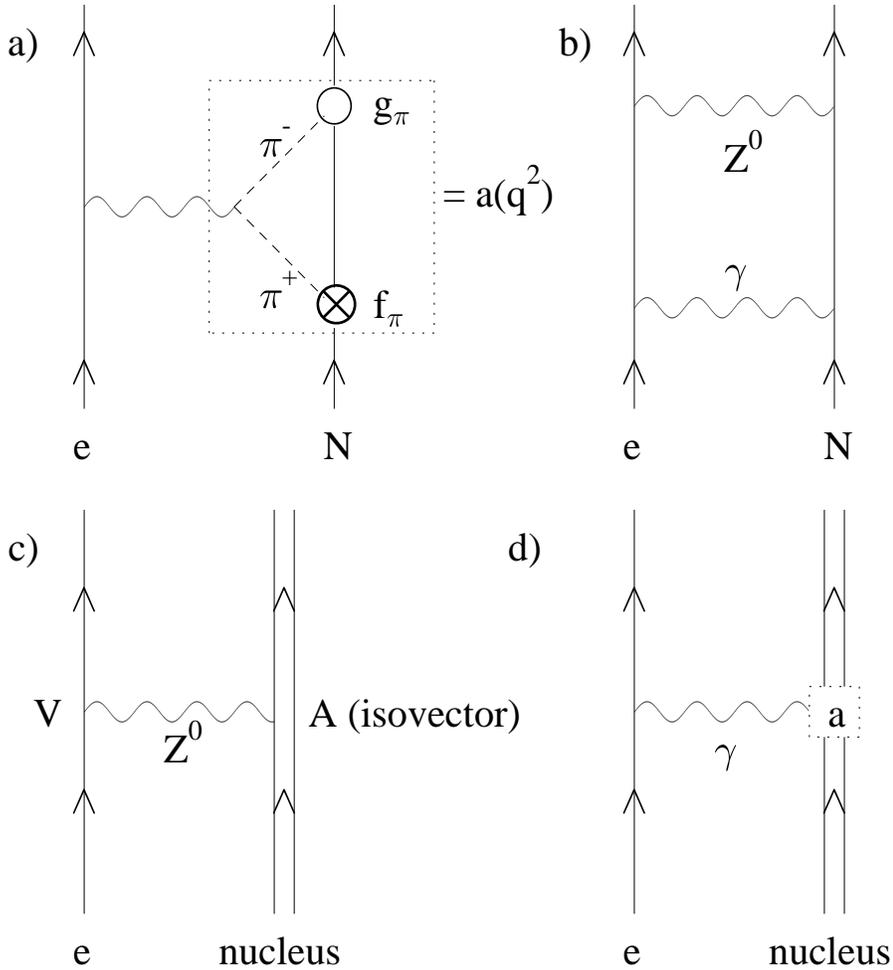,height=5.0in}
\caption{The nuclear anapole (a) and box (b) weak radiative 
corrections to the hydrogen atom.  The nuclear-spin-dependent
contribution to atomic PNC include the V(e)-A(nucleus) $Z^0$
exchange (c) and the electron interaction with the nuclear
anapole moment (d).}
\end{figure}
  
This raises two issues:
1) how does a composite object like a nucleus generate
an anapole moment? 2) how does the anapole contribution compare
to the tree-level V(e)-A(N) contribution? 
There are three separate contributions to the nuclear anapole moment: \\
1) The one-body contribution: 
\[ <g.s. | \sum_{i=1}^A (a_s + a_v \tau_3(i)) \vec{\sigma}(i)
|g.s.> \]
is a sum over the anapole moments of the individual nucleons.
As nucleons occupy spin-paired orbitals, this can be roughly
thought of as the anapole moment of the last, unpaired nucleon
(in the same sense that nuclear magnetic moments can be viewed
in this way).  If Fig. 4a is evaluated \cite{musolf} for a pion loop, one
finds that the isoscalar anapole moment is much larger than
the isovector, $a_s \gg a_v$. \\
2) The exchange-current contribution, consisting of diagrams
where the photon couples to a meson in flight between two 
nucleons, or to a nucleon-antinucleon pair. \\
3) The nuclear polarizability depicted in Fig. 5: the E1 
anapole operator couples the unperturbed ground state to the
opposite-parity components in the nuclear wave function that
arise from the hadronic weak interaction.  This third term
easily dominates the anapole moment of heavy nuclei and is
responsible for the anapole moment's $A^{2/3}$ growth, where
$A$ is the mass number.

\begin{figure}[htb]
\psfig{bbllx=-1.0cm,bblly=6.3cm,bburx=16cm,bbury=13.3cm,figure=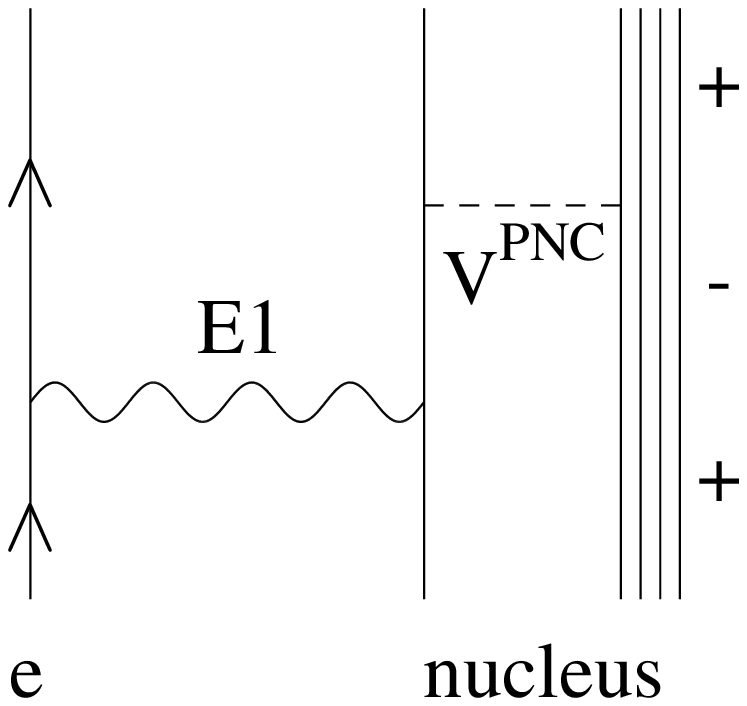,height=2.0in}
\caption{The PNC nuclear polarizability contribution to the
electron's interaction with the nuclear anapole moment.  This
diagram generally dominates the anapole moment of heavy nuclei.
The nuclear E1 operator is an unfamiliar one:
the usual operator vanishes due to the extended Siegert's 
theorem (see \cite{musolf}).  The evaluation of the 
polarizability requires a summation over a complete set of
intermediate nuclear states.  In the fully-interacting shell
model approach of \cite{musolf}, this sum was evaluated by
closure, assuming an average excitation energy
for the intermediate nuclear states.  Because of properties
of the E1 operator in the standard shell model space for
$^{133}$Cs (there are no nonzero matrix elements of J=1
odd-parity operators in the $1g_{7/2}2d_{5/2}3s_{1/2}2d_{3/2}1h_{11/2}$
space), the polarizability then
reduces to an expectation value of a two-body operator.}
\end{figure}
  
In the Cs experiment, 7000 hours of data taking yielded
the hyperfine-dependent (i.e., nuclear spin-dependent)
contribution to the atomic PNC signal.  The result
can be expressed as the strength of the nuclear-spin-dependent
electron-nucleus contact interaction
\[ H_W^{e-Nuc} \equiv {G_F \over \sqrt{2}} \kappa \vec{\alpha} \cdot
\vec{I} \rho(r) \] 
\[ \kappa = \underbrace{\kappa^Z_{V(e)-A(N)}}_{0.013} + \kappa_{anapole}
= \underbrace{0.112 \pm 0.016}_{7 \sigma} \]
The $Z^0$ contribution depends on the Weinberg angle (taken to be
$\sin^2\theta_W = 0.223$) and on the $^{133}$Cs ground state matrix element of the 
Gamow-Teller operator, which was taken from the shell model
calculation of \cite{musolf}.
The extraction of $\kappa$ from the Boulder result requires an
atomic calculation.  The above value is taken from Flambaum 
and Murray \cite{flambaum}.
As the extracted $\kappa$ is an order of magnitude larger that the
tree-level $Z^0$ contribution, the anapole contribution has been
clearly seen.  This places the following constraint on the weak
meson-nucleon coupling constants
\[ f_\pi - 0.21 (h_\rho^0 + 0.59h_\omega^0) = (0.99 \pm 0.16) \times
10^{-6} \]
where again a familiar combination of isoscalar weak couplings
arises.  (This result was calculated with
the same shell model techniques as in \cite{musolf}, but includes
the $\rho$ and $\omega$ exchange contributions to the nuclear
polarizability.  Note that the average nuclear excitation energy
was changed from the value used in \cite{musolf}, 15.2 MeV, to
9.3 MeV.)
In Fig. 6 this new constraint is shown along with those from three 
other PNC experiments ($^{18}$F, $^{19}$F, $\vec{p}+^4$He),
all of which test approximately the same combination of weak couplings.
It is immediately apparent that the $^{133}$Cs anapole moment 
is not in agreement with our tentative conclusion that $f_\pi$
is considerably below the DDH value.

\begin{figure}[htb]
\psfig{bbllx=-1.6cm,bblly=3.9cm,bburx=14cm,bbury=23.5cm,figure=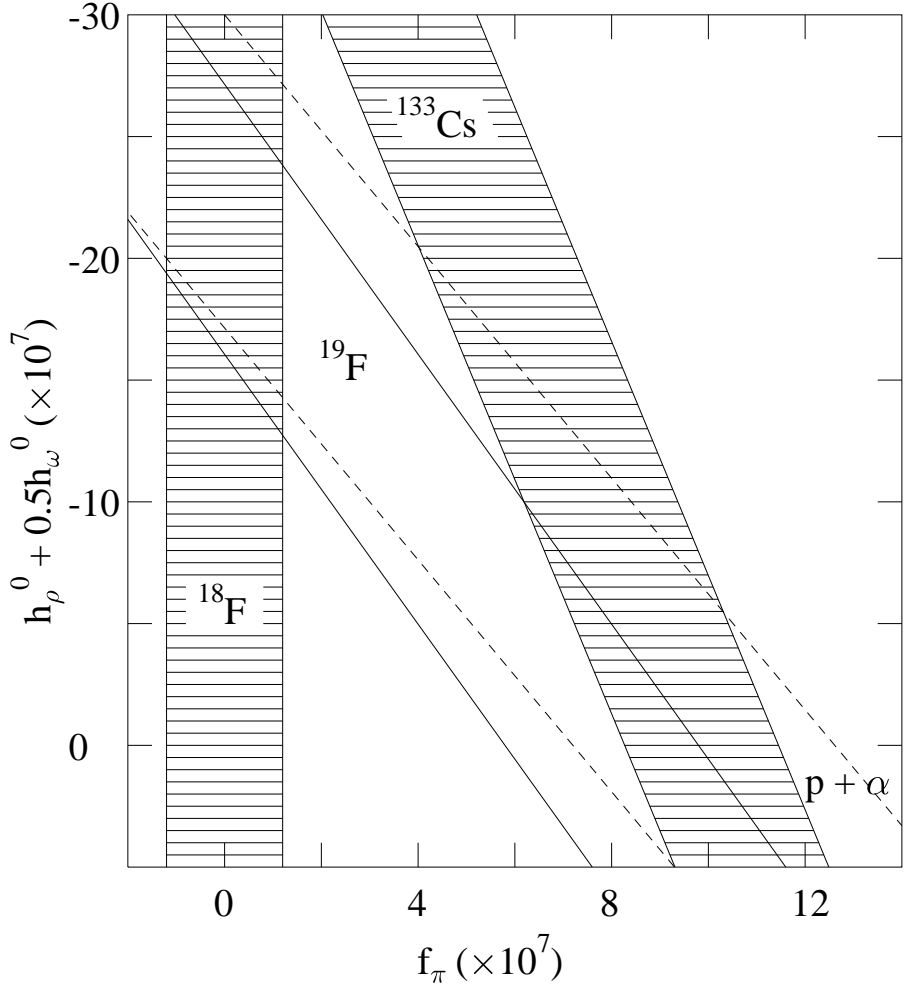,height=5.0in}
\caption{Experimental constraints on the isovector pion and 
isoscalar heavy meson weak couplings.  The indicated regions
correspond to the 1$\sigma$ experimental errors, only; no estimate
of the additional theory errors has been made.}
\end{figure}
  
Until the $^{133}$Cs anapole measurement there was very little 
redundancy among the experimental constraints shown in Fig. 6.
We now have to grapple with an inconsistency whose origin is
unknown.  It could be that one of the experiments is wrong.
The theory underlying Fig. 6 also has its weak points.  The
interpretation of the anapole moment measurement depends on
the accuracy of the calculation of the $^{133}$Cs polarizability.
The evidence that this has been done well is the reasonable agreement
between two independent calculations \cite{musolf,flambaum2},
but this should be further explored.  There are additional 
issues of concern --- the use of bare operators in the nuclear
calculations, strange quark contributions to the hadronic
matrix elements, etc. --- that require more technical discussion
than is possible here. 
   
There are several possibilities for new experiments that could
greatly clarify matters.  Perhaps the most important are 
possibilities for measuring the np weak interaction either
directly or in a few-body system that can be reliable 
interpretted.
Efforts are underway to measure the PNC $\vec{\mathrm{n}}$ spin rotation in $^4$He \cite{UW}, 
and a proposal has been made to measure A$_\gamma$ for $\vec{\mathrm{n}}$ + p $\rightarrow$ d + $\gamma$ \cite{LANL}.
There is also an important experiment nearing completion at TRIUMF
in which A$_L$ for $\vec{\mathrm{p}}$ + p is being measured
at medium energies, thereby testing the combination of vector
meson PNC couplings that contribute to p-d wave interference \cite{TRIUMF}. 
Perhaps our picture of hadronic PNC will be clearer at the time 
of INPC2001. 

\section{Atomic Nuclei and T Violation}
Recall our table of static electromagnetc couplings: 
\[ \begin{array}{cccc}
&\underline{CJ}&\underline{MJ}&\underline{EJ} \\
J=0& PT&& \\
J=1 & \not P \not T & PT & \not P T \\
\vdots& \vdots& \vdots& \vdots 
\end{array} \]
The moment of interest in this section is the $C1$, the P-odd
T-odd electric dipole moment.  The required CP-violating, P-violating
interactions occur in nature ($K_L \rightarrow \pi\pi$) and
in the standard model (through the CKM phase in the quark mass
matrix and the $\bar{\theta}$ term), and there are strong reasons for
believing additional sources of P-violating, CP-violating interactions
exist beyond the standard model.  Such interactions will generate
a nucleon edm:
\[<p'|J_\mu^{em}|p> = \ldots + \bar{N}(p') \underbrace{d(q^2)}_
{edm} \sigma_{\mu \nu} q^\nu \gamma_5 N(p) \]
  
If a neutron or neutral atom having an edm --- a charge separation
along the direction of spin --- is placed in an external electric
$\vec{E}_{ext} = E_z \hat{z}$, the interaction is
\[ H = d \vec{s} \cdot \vec{E} \]
The resulting torque makes the spin precess about $\hat{z}$.
Neutron measurements by the Grenoble \cite{grenoble} and Gatchina
\cite{gatchina} groups and atomic measurements by the Seattle
group \cite{seattle} have produced wonderfully precise results
\[ |d_{\mathrm{neutron}}| \leq 8 \cdot 10^{-26} \mathrm{e~cm} \]
\[ |d(^{199}\mathrm{Hg})| \leq 1.3 \cdot 10^{-27} \mathrm{e~cm} \]
It is the rapid improvement in the atomic measurements that 
motivates my inclusion of this topic here.  In many cases
the most interesting sources of CP violation within the atom 
are those residing within the nucleus.  Thus we must deal with
the nuclear physics governing CP violation to understand
the implications of atomic edm measurements for underlying
theories of CP violation.  This involves a series of steps:
understanding how the underlying model generates a nucleon edm
and a CP-violating NN interaction; calculating how this 
interaction polarizes the nucleus to produce a nuclear edm;
and calculating the atomic polarization that the nuclear edm
induces. 

\begin{figure}[htb]
\psfig{bbllx=0.0cm,bblly=13.7cm,bburx=18cm,bbury=21.8cm,figure=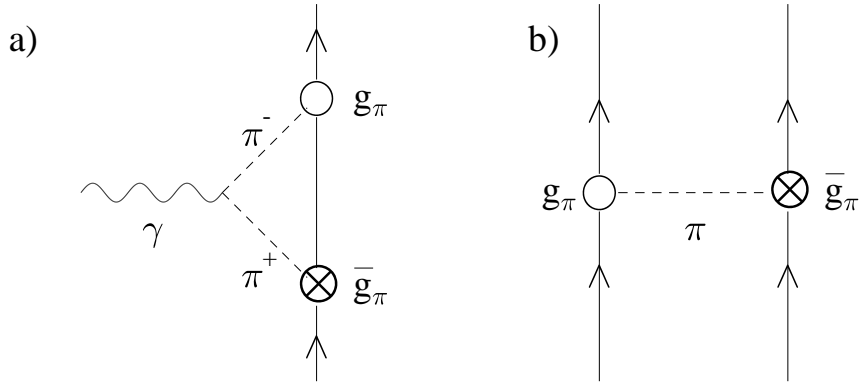,height=2.0in}
\caption{The nucleon edm (a) and P-violating, CP-violating NN
interaction (b) that will arise from a CP-violating scalar coupling
$\bar{g}$ of the pion to the nucleon.}
\end{figure}
  
Neutrons and atomic nuclei also provide complementary constraints
on CP-violation.  Consider, for example, the $\bar{\theta}$ term in
the QCD Lagrangian
\[ \bar{\theta} {g^2 \over 32 \pi^2} F_{\mu \nu} \tilde{F}^{\mu \nu} \]
where the neutron edm limits $|\bar{\theta}| \lsim 10^{-10}$.
The unnatural smallness of $\bar{\theta}$ is often called the strong
CP problem.  One low-energy consequence of $\bar{\theta}$ is the
generation of a CP-violating scalar coupling $\bar{g}_{\pi N N}$
of the pion to the nucleon \cite{crewther}
\[  {L}_{\pi NN} \rightarrow
\vec{\pi} \cdot \bar{N} \vec{\tau} (i \gamma_5 g_{\pi NN} +
\bar{g}_{\pi NN}) N \]
\[ \bar{g}_{\pi NN} \sim 0.03 \bar{\theta} \]
  
The pion coupling is important because the pion is the longest
range meson and thus can produce the largest charge separation,
or edm, as shown in Fig. 7a.  It also should dominate the 
long-range CP-violating NN interaction (Fig. 7b).  The neutron
edm corresponding to Fig. 7a is \cite{crewther}
\[ d_n \sim {e g_{\pi NN} \bar{g}_{\pi NN} \over 4 \pi^2 M} 
ln({M \over m_\pi}) \]
where we note the chiral log.  Thus an atomic edm measurement
could be analyzed in terms of $\bar{g}_{\pi N N}$, which then serves
as a low-energy constraint on possible models of CP violation. 
  
The isospin dependence of the long-range $\pi N N$ interaction
can distinguish competing descriptions of CP-violation
\cite{hhenley,herczeg}.  The possibilities include isoscalar,
isovector, and isotensor couplings
\[ \begin{array}{l}
\bar{g}^0_{\pi NN} \bar{N} \vec{\tau} N \cdot \vec{\pi}  \\
\bar{g}^1_{\pi NN} \bar{N}N \pi^0  \\
\bar{g}^2_{\pi NN} \bar{N}(3 \tau_3 \pi^0 - \vec{\tau} \cdot \vec{\pi})N \end{array} \]
where $\bar{\theta}$ generates the first (isoscalar) coupling
above.  The neutron edm limit provides the constraint
\[ |d_n^{exp}| \lsim 8 \cdot 10^{-26} \mathrm{e~cm} \Rightarrow
|\bar{\theta}| \lsim 10^{-10} \]
A calculation shows that the $^{199}$Hg edm would arise primarily
from the CP-odd mixture in the ground state due to the CP-violating
NN interaction, rather than from the edm of the unpaired valence
nucleon.  With some effort the constraint imposed by the Seattle
experiment can be recast in the form
\[ |d_n^{\bar{\theta}}| \lsim 5 \cdot 10^{-26} \mathrm{e~cm} \]
There are several points to be made.  First, the atomic and
neutron edm measurements provide very similar constraints
on $\bar{\theta}$.  Second, the atomic limit is mildly dependent
on the souce of the CP violation: it tightens
by about a factor of four if the CKM phase is used as the
source of the CP violation.  Third, the isospin dependence of
$\bar{g}$ leads to a distinctive scaling of the atomic
result with the N and Z of the nucleus.  In the case of
the isoscalar coupling (i.e., $\bar{\theta}$), the atomic
edm $\sim$ (N-Z).  Thus if a nonzero neutron edm were measured,
one could do additional atomic experiments looking for such
a dependence, in order to clarify the origin of the neutron
result. 

With the recent rapid progress in atomic edm measurements, this 
approach now compares favorable to neutron edm experiments in
sensitivity to the underlying particle physics.
But the progress is not at an end: both types of experiments 
continue to improve, with current efforts possibly yielding
another factor of 20.  In the case of the atomic experiments
the detailed dependence of the resulting
limits on the nuclear physics --- illustrated here for $\bar{\theta}$ ---
points out the relevance of our field to this endeavor.
The accuracy of the resulting particle physics constraints will
depend on the quality of the supporting nuclear calculations.  Finally, there is the exciting
possibility that the enhancements we have long exploited
in PNC tests might find an analog in atomic edm measurements.
To date none of the nuclei identified as especially polarizable
(some by enhancements factors ranging up to 10$^4$ \cite{hhenley})
have been suitable candidates for atomic measurements.
But as new atomic methods are developed, this situation could
certainly change.
  
\section{$\beta$ Decay Tests of Weak Interactions}
As I mentioned in the introduction, $\beta$ decay studies played
a prominent role in establishing the experimental foundations
of the standard model and, today, continue to be important as
tests of new physics.  There is a great deal of new activity,
much of it reported in the parallel sessions of this meeting.
While I no not have the time to discuss any of this in the
detail it deserves, I do want to cite a few of the results of
the past year that I found notable:\\
$\bullet$ Adelberger is reporting at this meeting a new study
of the e$^+-\nu_e$ correlation in $0^+ \rightarrow 0^+ \beta$
decay which tightens constraints on scalar interactions.
Quoted in terms of an equivalent boson mass, the result
is $M_S \gsim 4 M_W$. \\
$\bullet$ Savard is reporting on the status of Fermi $\beta$
decay tests of standard model unitarity.  While there is still
a modest departure from unitarity, the data for targets with
$5 \lsim \mathrm{Z} \lsim 26$ smoothly extrapolate to small
Z.  Perhaps this indicates that calculations of the weak
radiative corrections and effects of isospin mixing are 
in reasonably good shape. \\
$\bullet$ Tritium $\beta$ decay constraints on the $\nu_e$ mass
continue to improve, with the Mainz and Troitsk experiments 
yielding constraints $\lsim (3-5)$ eV.  Some progress has been
made in identifying effects responsible for a troubling excess
of electrons very near the endpoint (the so-called negative
$m_\nu^2$ problem).  Yet it is clear the spectrum shape is not
yet fully understood \cite{nu98}. \\
$\bullet$ Atomic exchange effects --- where the emitted $\beta$
ray is captured into an atomic orbit while an atomic electron
is kicked into the continuum --- were recently measured for
the first time in an allowed decay \cite{angrave}. \\
$\bullet$ Environmental effects in $\beta$ decay --- effects
of the surrounding atoms of the spectrum of emitted electrons ---
have been seen for the first time \cite{gatti}.  Such effects
were predicted some time ago \cite{koonin}. 

\section{Neutrino Masses}
I would like to close this talk with a few comments about neutrino
mass, given the excitement this field has generated these past
few months.  It has long been realized that neutrinos might 
provide a special window on physics far beyond the standard model.
This became apparent from a puzzle that arose when extended
models were first considered.  In such a model one might hope
to replace the standard model doublets with larger multiplets
in the hope of further unifiying the interactions, as depicted
below
\[ \left( \begin{array}{c} \nu \\ e \end{array} \right)
\begin{array}{c} e.g. \\ \longrightarrow \\ grander \\ model
\end{array} \left( \begin{array}{c} u \\ d \\ \nu \\ e \end{array}
\right)  \]
The puzzle derives from the masses of the particles in the 
multiplet.  One would expect these particles to couple to the
mass-generating fields in a similar way, so that the resulting
masses would be comparable up to group theory factors.
But, while the electron and first generation quarks have masses
on the order of an MeV, the $\nu_e$ mass is at least six orders
of magnitude smaller. 

Gell-Mann, Ramond, Slansky, and Yanagida \cite{seesaw} recognized
that the neutrino is special.  Unlike the other fermions which
clearly have distinct antiparticles under particle-antiparticle
conjugation (e.g., $e^- \rightarrow e^+$, with
the electron and positron distinguished by their opposite charges),
there is no obvious additive quantum number distinguishing the
$\nu$ from the $\bar{\nu}$.
This has the consequence that in addition to the usual Dirac mass
term $m_D$, neutrinos can have Majorana masses that break lepton
number conservation.  The result is a neutrino mass matrix
that, when diagonalized, yields a light neutrino mass
\[ m^{light}_{\nu} \sim m_D \underbrace{\left( {m_D \over M_R} \right)}_
{``small~parameter''} \]
where $M_R$ is some heavy right-handed Majorana mass characterizing
scales beyond the standard model.  If one fixes $m_D$ to
the charged fermion masses and $m_\nu^{light}$ to theoretical
scenarios explaining the solar and atmospheric neutrino problems,
then
\[ M_R \sim (10^{12} - 10^{16}) \mathrm{GeV} \]
This ``seesaw" explanation of light neutrino masses suggests
that neutrinos provide a window on new physics far beyond the
explored low-energy world of the standard model. 

To provide some picture of how these various results might fit
together to form some pattern, I now discuss a recent 
paper by Georgi and Glashow \cite{georgi98}.  The assumptions
of their construction are:\\
$\bullet$ Three light Majorana neutrinos \\ 
$\bullet$ The atmospheric neutrino problem is due to 
$\nu_\mu \rightarrow \nu_\tau$ oscillations, since the 
$\nu_\mu \rightarrow \nu_e$ alternative is ruled out by the
Chooz experiment. \\
$\bullet$This oscillation is nearly maximal with 
sin2$\theta_{23} \sim$ 1 and
5 $\cdot 10^{-4}$ eV$^2 \lsim \delta m_{23} \lsim 6 \cdot 10^{-3}$ eV$^2$. \\
$\bullet$The solar neutrino problem is due to oscillations with
$6 \cdot 10^{-11}$ eV$^2 \lsim \delta m^2 \lsim 2 \cdot 10^{-5}$ eV$^2$. \\
$\bullet$ The neutrino masses are constrained to satisfy
m$_1$+m$_2$+m$_3 \sim$ 6 eV in order to generate hot dark 
matter for large scale structure formation (a somewhat 
speculative condition). \\
$\bullet$The absence of neutrinoless double $\beta$ decay requires  
$\langle \mathrm{m}^{Maj}_\nu \rangle \lsim 0.4 \mathrm{eV}$,
so choose $\langle \mathrm{m}^{Maj}_\nu \rangle \sim$ 0. \\ 
$\bullet$Because of the LSND/KARMEN conflict, the LSND results 
are not considered. 
  
These constraints lead to a pattern of three nearly degenerate
massive neutrinos with m$_i \sim M$ and a simple mass matrix
that accounts for the atmospheric and solar neutrino problems
through vacuum oscillations,
\[  M \left( \begin{array}{ccc}
0 & {1 \over \sqrt{2}} & {1 \over \sqrt{2}} \\
{1 \over \sqrt{2}} & {1 \over 2} & {-1 \over 2} \\
{1 \over \sqrt{2}} & {-1 \over 2} & {1 \over 2} \end{array} \right)
\begin{array}{c} \nu_e \\ \nu_\mu \\ \nu_\tau \end{array} \]
This kind of mass matrix can arise naturally in model schemes,
as has been shown recently by Mohapatra and Nussinov \cite{mohapatra}.  Clearly it is just
one possibility among many, but suggests that the hints of
massive neutrinos we now have may yet conform to a simple 
pattern. 

The interest in neutrino physics is likely to build in the next 
few years, with nuclear physics having many opportunities to
contribute.  The hope is that we will soon understand whether 
a simple pattern like that given above describes nature.  So
in my wishlist for our field in the years leading up to INPC2001
I would include \\
$\bullet$ LSND vs. KARMEN fully resolved \\
$\bullet$ data from SNO \\
$\bullet$ improved $\beta \beta$ decay limits reaching well beyond
0.1 eV \\
$\bullet$ some progress on very long baseline $\nu_\mu \leftrightarrow
\nu_\tau$ experiments. \\
There is a great deal new to anticipate. 

This work was supported in part by the US Department of Energy.


\begin{thebibliography}{9}
\bibitem{adel} E. G. Adelberger and W. C. Haxton,
               Ann. Rev. Nucl. Part. Sci. 35 (1985) 501.
\bibitem{ddh} B. Desplanques, J. F. Donoghue, and B. R. Holstein,
              Ann. Phys. (NY) 124 (1980) 449.
\bibitem{boulder} C. S. Wood et al., Science 275 (1997) 1759.
\bibitem{zeld} Ya. B. Zeldovich, Sov. Phys. JETP 6 (1958) 1184
               and references therein.
\bibitem{khrip} For early arguments that the anapole moment
                might be measurable in a heavy nucleus, see
                V. V. Flambaum and I. B. Khriplovich, Sov.
                Phys. JETP 52 (1980) 835.
\bibitem{musolf} W. C. Haxton, E. M. Henley, and M. J. Musolf,
                 Phys. Rev. Lett. 63 (1989) 949.
\bibitem{flambaum} V. V. Flambaum and D. W. Murray, Phys.
                   Rev. C56 (1997) 1357.
\bibitem{flambaum2} V. V. Flambaum, I. B. Khriplovich, and O. P.
                    Sushkov, Phys. Lett. B146 (1984) 367.
\bibitem{UW} E. G. Adelberger and B. Heckel, private communication.
\bibitem{LANL} J. D. Bowman, private communication.
\bibitem{TRIUMF} A. R. Bedoz et al., Nucl. Phys. A629 (1998) 433c.
\bibitem{grenoble} K. F. Smith et al., Phys. Lett. B234 (1990) 191.
\bibitem{gatchina} I. S. Alterev et al., Phys. Lett. B276 (1992) 242.
\bibitem{seattle} J. P. Jacobs et al., Phys. Rev. Lett. 71 (1993) 3782.
\bibitem{crewther} R. J. Crewther, P. DiVecchia, G. Veneziano,
                   and E. Witten, Phys. Lett. 88B (1979) 123 and
                   91B (1980) 487.
\bibitem{hhenley} W. C. Haxton and E, M. Henley, Phys. Rev. Lett.
                  51 (1983) 1937.
\bibitem{herczeg} P. Herczeg, Hyp. Int. 43 (1988) 77.
\bibitem{nu98} C. Weinheimer and V. M. Lobashev, talks presented at
               Neutrino '98 (June, 1998, Takayama, Japan).
\bibitem{angrave} L. C. Angrave et al., Phys. Rev. Lett. 80 (1998) 1610.
\bibitem{gatti} F. Gatti et al., Genoa Univ. preprint (1998).
\bibitem{koonin} S. E. Koonin, Nature 354 (1991) 468.
\bibitem{seesaw} M. Gell-Mann, P. Ramond, and R. Slansky, in
                 Supergravity, eds. P. Van Nieuwenhuizen and D. Z.
                 Freedman (North Holland, Amsterdam, 1979) p. 315;
                 T. Yanagida, Proc. of the Workshop on Unified
                 Theory and Baryon Number in the Universe, eds.
                 O. Sawada and A. Sugamoto (KEK, 1979).
\bibitem{georgi98} H. Georgi and S. L. Glashow, hep-ph/9808293.
\bibitem{mohapatra} R. N. Mohapatra and S. Nussinov, hep-ph/9809415v2.
\end{thebibliography}
\end{document}